\documentclass[12pt]{article}
\usepackage{amsmath}
\usepackage{graphicx}
\usepackage{enumerate}
\usepackage{longtable}
\usepackage{multirow}
\usepackage{siunitx}        
\usepackage{xcolor}
\usepackage{natbib}
\usepackage{url} 

\newcommand{\blind}{1}
\def\mbf#1{\mathbf{#1}}
\def\ul#1{\underline{#1}}

\begin{document}

\def\spacingset#1{\renewcommand{\baselinestretch}%
{#1}\small\normalsize} \spacingset{1}

\def\ourtitle{Estimating Sleep \& Work Hours from Alternative Data by Segmented Functional Classification Analysis (SFCA)}

\if1\blind
{
  \title{\bf \ourtitle}
  \author{Klaus Ackermann
  \\
    {\small Dept. of Econometrics \& Business Statistics, and}\\
    {\small SoDa Laboratories, Monash Business School, Monash University}\\
    Simon D Angus \\
    {\small Dept. of Economics, and}\\
    {\small SoDa Laboratories, Monash Business School, Monash University}\\
    and \\
    Paul A Raschky \\
    {\small Dept. of Economics, and}\\
    {\small SoDa Laboratories, Monash Business School, Monash University}}
  \maketitle
} \fi

\if0\blind
{
  \bigskip
  \bigskip
  \bigskip
  \begin{center}
    {\LARGE\bf \ourtitle}
\end{center}
  \medskip
} \fi

\begin{abstract}
\noindent Alternative data is increasingly adapted to predict human and economic behaviour. This paper introduces a new type of alternative data by re-conceptualising the internet as a data-driven insights platform at global scale. Using data from a unique internet activity and location dataset drawn from over 1.5 trillion observations of end-user internet connections, we construct a functional dataset covering over 1,600 cities during a 7 year period with temporal resolution of just 15min. To predict accurate temporal patterns of sleep and work activity from this data-set, we develop a new technique, Segmented Functional Classification Analysis (SFCA), and compare its performance to a wide array of linear, functional, and classification methods. To confirm the wider applicability of SFCA, in a second application we predict sleep and work activity using SFCA from US city-wide electricity demand functional data. Across both problems, SFCA is shown to out-perform current methods.

\end{abstract}
\noindent%
\vfill
\newpage
\spacingset{1.5} 



\section{Introduction}\label{sec:intro}

At what time of day will road accidents on a major freeway most likely occur given that historical data on driver demographics, vehicle type, usage, and weather are available at hourly resolution? At what age is a particular degenerative disease most likely to strike an individual given that a large number of body mass, height, morbidity and diet composition trajectories by age are available for a comparison sample?

Both of these questions, and thousands like them, are well handled with the machinery of functional data analysis~\citep{Ramsay:2006vf} where either outcomes or predictors are given in the form of trajectories. In the questions posed above, and the focus of this paper, a functional or scalar outcome, $y$ is associated with one or more predictor trajectories of the kind $x(t)$. In the language of functional data analysis, these questions conform to the \emph{scalar-on-function} class of problems ~\citep{Reiss:2016fj}. In this literature, $x(t)$ is obtained by some smoothing or gap-filling procedure from sampled predictor observations $(t_i,x_i)$. The problem is then one of estimating or learning the mapping $f: x(t)\mapsto y$. With a set of observations on $y$, and trajectories $x(t)$, a standard approach would be to apply Functional Principal Component Analysis (FPCA)~\citep{Silverman:1996km,Shang:2014gm} or its variants~\citep{Reiss:2007hv,Yao:2005he,Chen:2018hq} which, after first applying PCA to the predictors $x(t)$ for dimension reduction, applies a regression framework to establish an estimate of $f$.

However, the two questions posed above share a key property that we wish to exploit in the current study: the domain of $y$, $\mathbf{Y}$ is coincident with the domain of $t$, $\mathbf{T}$. Or in scalar-on-function regression terms, the scalar is in the domain of the predictor function. The first question is an example of the wide class of problems where the outcome (`When') is in the temporal domain (`time') over which predictor trajectories are described, and the second poses an outcome (`What age') which is associated with predictor trajectories over subject ages. Indeed, any ordered domain which both supports the predictor trajectories and contains the outcomes will be applicable, i.e. where $\mathbf{Y} \subseteq \mathbf{T}$. These data need not arise from individual trajectories across sub-domains of $T$ but could, as with any functional data analytic problem, arise from independent observations at one or more locations in $\mathbf{T}$, which are then composed into the object $x(t)$ for analysis.

Here we demonstrate that for problems fitting the general characterisation,
\begin{equation}\label{eq:main}
    f: \mathbf{X} \times \mathbf{T} \rightarrow \mathbf{Y}, \,\, \mathrm{where}\,\, \mathbf{Y} \subseteq \mathbf{T}
\end{equation}
an alternative approach to the estimation of $f$, which we call \emph{Segmented Functional Classification Analysis} (SFCA), is both easily implemented in standard statistical software, and out-performs a variety of functional and non-functional methods on two exemplar applications.


Our first application, predicting sleep and work start-time, stop-time and duration from geo-located internet activity data, additionally introduces an entirely novel and rich internet activity dataset which arises from over 1 trillion unique observations of online/offline internet connectivity globally during 2006-2012. These data, when accurately geo-located and aggregated produce 24 billion observations of the fraction of online internet addresses found at any one of 1,647 urban-boundaries, during 15min intervals, forming rich predictor trajectories from which to estimate offline, temporally identified, activity classes as outcomes. Using American Time Use Survey (ATUS) population-weighted averages of start-time and stop-time of activity categories `sleep' and `work' as outcomes, the SFCA approach introduced in this paper, achieves the lowest cross-validated RMSE of 17 modelling approaches from across the functional regression, regression, penalised regression, and classification tool-sets. In performance terms, the best SFCA method (random forest classification) gains between 37\% and 68\% improvement in (geometric-)average RMSE when directly compared with standard functional regression methods on either sleep, or work, start- and stop- time estimation.


Our second application, predicting the same ATUS activity category classes as outcomes from US electricity demand data, confirms the performance of SFCA relative to all other methods. We geo-match Federal Electricity Regulatory Commission (FERC) Form 714 demand data at electricity balancing authority regions to ATUS city boundaries, down-scaling the hourly reporting data to a range of 15min interval predictor trajectories. Once again, the best SFCA method (either random forest or boosted trees) achieves between 4\% and 48\% average performance (RMSE) gain over the best performing functional data analysis method.


SFCA re-conceptualises the estimation problem in classification terms, exploiting the property of this class of problems, $\mathbf{Y} \subseteq \mathbf{T}$, by applying the standard, variously named, `stack', `pack', or `wide to tall'  algorithm, such that all trajectories, $x(t)$ and outcomes, $y$ are transformed to standardised segments of $\mathbf{T}$. In so doing, the powerful and rapidly developing technology of \emph{classification} can be employed. We attribute the strong performance of the SFCA method in these exemplar problems to the known ability of classification methods to learn specific, informative features from complex, high-dimensional data. 

SFCA also over-comes the curse of dimensionality which arises in most functional data analytic problems necessitating some form of dimension reduction to succeed.  Whereas a typical functional data analytic problem may constitute $N$ outcome observations associated with $M$ trajectory dimensions of length $T$, producing an $N \times MT$ problem, as will be shown, SFCA dramatically transforms the problem into shape $NT \times M$, meaning that no dimensional reduction is required and all information can be used to learn $f$. Indeed, the stacking step of SFCA encourages the addition of new predictors via feature engineering, as is standard in statistical machine learning, increasing the ability of SFCA to identify more precisely the particular location(s) in $\mathbf{T}$ of the outcome label.


Our paper contributes to three broad strands of literature: First, the method we introduce, SFCA, can be considered as a methodological bridge between two important fields: functional data analysis~\citep{Ramsay:2006vf} on the one hand, and statistical machine learning~\citep{hastie:2009} on the other. By re-casting functional data analysis problems into classification problems, a wide array of applications which fit the SFCA inclusion criterion will likely benefit from the SFCA machinery demonstrated here.

Second, this paper introduces a novel granular dataset on internet activity to the rapidly growing literature that applies passively collected `big data' to the pursuit of quantitative social science. Prominent sources of such data include cell-phone meta-data~~\citep{Onnela:2007ek,Blumenstock:2015,Toole:2012jh}, ISPs~~\citep{BenitezBaleato:2015hk, Weidmann:2016jv},  `app' activity logs~~\citep{Ginsberg:2009ep,Franca:2015dv,Bakshy:2015jn}, or night-time satellite imagery~~\citep{Jean:2016,Chen:2011,Henderson:2012,Raschky:2014vm}. 

Within this literature, our paper is most closely related to studies using geolocated Internet and communication technology (ICT) data such as \citet{Blumenstock:2015} \citet{BenitezBaleato:2015hk} and \citet{Weidmann:2016jv}. In contrast to existing work, our paper introduces a dataset that features consistently measured, intra-diurnal observations at a global scale. Our data builds a complement to the existing alternative datasets that allows one to predict intra-diurnal, human behaviour at a global scale.

Third, we also speak to the large literature on the 
prediction of electricity demand \emph{from} human behavioural-, economic- or environmental- signals~\citep{Hyndman:2010co, Hendricks:2012hz, Cabrera:2017be, BenTaieb:2020hj, An:2013cs, Muratori:2013dc, Bogomolov:2016et}. We expand this this literature by conducting a reverse exercise and demonstrate that intra-diurnal electricity demand data can be used to \emph{predict} human behavioural patterns.

\section{Segmented Functional Classification Analysis (SFCA)}

The key inclusion criteria for applying SFCA to a functional data analytic problem is given in \eqref{eq:main}, namely that $\mathbf{Y} \subseteq \mathbf{T}$. With the domain criteria satisfied, one can then proceed to apply the SFCA algorithm.

In general terms, the SFCA algorithm comprises four fundamental steps:
\begin{enumerate}
    \item {\bf Pre-process}: Pre-process predictor trajectories (up- or down- scaling, interpolation etc.) such that each observation of $x_m(t)$ for predictor dimensions $m\in\{1\dots M\}$ over $N$ independent observations is defined over a set of consistently spaced, discretised \emph{segments}, $s \in \{1,\dots,S\}$, giving $x_m(s)$ of size $[N \times MS]$.
    \item {\bf Stack}: Apply the stack (or `pack', or `wide-to-tall') algorithm to $x_m(s)$, transposing each predictor trajectory over the segment domain from $1 \times S$ to $S \times 1$ giving a new object, $\mathbf{x}$ of size $[NS \times M]$
    \item {\bf Threshold}: Define a threshold criteria for outcomes such that each observation of $y$ is converted to a balanced boolean vector, $\mathbf{y}$ over the same discretised segment domain $s \in \{1,...,S\}$ (size $[NS \times 1]$)
    \item {\bf Estimate / Learn}: Estimate, or learn, the mapping $f : \mathbf{x} \mapsto \mathbf{y}$ with a preferred tool.
\end{enumerate}

\subsection{Considerations}

\paragraph{Granularity} When pre-processing, it is important to keep in mind that the choice of the discretisation regime $\{1,\dots,S\}$ will influence the precision with which the thresholded rendition of outcomes can be applied, and so, may influence the overall precision of the final model. For instance, if working with the temporal domain, an outcome given with precision of seconds may imply defining a segmentation regime at seconds precision over $\mathbf{T}$. However, the granularity decision will most heavily be informed by the quality of the predictor observations and their consistency. If the predictor trajectories have been observed at hourly granularity, down-scaling to seconds will expose the approach to strong reliance on down-scaling methods, which may introduce unwanted variation, potentially confusing down-stream estimation and learning.  On the other hand, one can add smoothing, filtering or other methods to the learning loop to overcome implied coarseness in segmentation of $\mathbf{T}$.

\paragraph{Feature engineering} In practice, as with other functional data analytic methods (e.g. \cite{Reiss:2007hv}) one typically builds a range of derivative features from the set of original predictor trajectory observations (e.g. first moment, second moment, etc.). In applying SFCA, the same applies with the added opportunity that derivative features can be built from columns of the stacked matrix $\mathbf{x}$. However, although building features at this later stage is perhaps a more natural approach for practictioners of machine learning (`feature engineering'), caution is needed in SFCA due to the stack step. It is crucial that one keeps track of the beginning and end of each segmented trajectory series so that new features are not composed across $\mathbf{T}$ domain boundaries. Nonetheless, the region of potential features that can be created is often vast, with moments, lags, left- or right- shifts, or combinations thereof all sensible approaches to retain the key features of the predictor trajectories in the transformed problem.

\paragraph{Thresholding \& label balance} Typically, for reasons of balance, encoding a single element in $\mathbf{y}$ as class 1 and all other elements 0 will be an unsuccessful approach with classification technology.  However, applying a simple threshold or defining a small range of temporal segments around $y_i$ as class 1 are examples of simple options to produce a well balanced classification target under many situations. For scalar outcomes $y$, a facile way to build $\mbf{y}$ is simply to leverage the inclusion criteria $\mbf{S} \subseteq \mbf{T}$ and apply a binary decision rule,
\[
\mbf{y}_{i}
= \begin{bmatrix}
           \mbf{y}_{i,1} \\
           \vdots \\
           \mbf{y}_{i,S}
\end{bmatrix}
= 
\begin{cases}
  1 & \text{for} \quad s \leq  y_i \\
  0 & \text{otherwise}\, ,
\end{cases}
\]
defining the classification-ready outcome for outcome $y_i$ over $S$ segments. So long as the outcome has sufficient variance over the segmentation interval $1\dots S$, a balanced label vector $\mbf{y}$ is obtainable. Where this is not the case, narrowing the segmentation window can recover balance, with feature engineering around shifts and lags often useful to ensure no important predictor trajectory features are lost by narrowing the segmentation window.

\subsection{Example}

Suppose that $n=2$ observations have been made of outcomes, $y_1 = 9$, and $y_2 = 13$, and corresponding predictor trajectories, $x_{i,m}$ for each observation $i$ and dimension $m$ as given in Table~\ref{tab:example} below and we wish to apply SFCA to these data.

\begin{table}[htbp]
\caption{Example functional data}\label{tab:example}
\centering
\begin{tabular}{cccccccccc}
\hline
\hline
\multirow{2}{*}{$y$} &
    \multirow{2}{*}{$x_{i,m}$} &
    \multicolumn{8}{c}{$t$} \\
\cline{3-10}
    & & 7 & 8 & 9 & 10 & 11 & 12 & 13& 14 \\
\hline
\multirow{2}{*}{9}  & $x_{1,1}$ & 1 & 0 & 1 & 3 & 5 & 3 & 2 & 1 \\
                    & $x_{1,2}$ & 1 & 3 & 7 & 5 & 2 & 1 & 0 & 1 \\
\multirow{2}{*}{13}  & $x_{2,1}$ & -3 & -4 & -2 & -1 & 0 & 1 & 3 & 4 \\
                    & $x_{2,2}$ & 12 & 13 & 14 & 15 & 16 & 16 & 16 & 16 \\
\hline                    
\end{tabular}
\end{table}

Pre-processing the trajectory data in this case is not necessary, as we have a complete data-set across each trajectory time-point, dimension and observation. Segmentation can be defined simply by re-labelling $t_1$ to $s=1$, $t_2$ to $s=2$, and so on. By dimensional concatenation, we have $x_m(s)$ of size $[2 \times (2*8)]$, a `wide' array.

Stacking the wide array into a `tall' array is accomplished by applying a transpose to each dimension, independently. For standard statistical packages, the same is typically achieved by providing a compressed wide table (such as Table~\ref{tab:example}) to the function, the variable names over which stacking should occur (e.g. \verb+t_7+, \verb+t_8+, etc.), and an indicator variable which contains the dimensional labels (e.g. \verb+x_im+). The result is a tall array (size $[16 \times 2]$),
\[
\mbf{x}
= \begin{bmatrix}
           1 & 1 \\
           0 & 3 \\
           \multicolumn{2}{c}{$\vdots$} \\
           2 & 0 \\
           1 & 1 \\
           -3 & 12 \\
           -4 & 13 \\
           \multicolumn{2}{c}{$\vdots$} \\
           3 & 16 \\
           4 & 16 \\
\end{bmatrix}
\]

Next, we build $\mbf{y}$ via thresholding. Here we apply the simple less than / greater than rule mentioned earlier to build two, stacked, boolean vectors corresponding to the outcome. Since $y_1 = 9$, and $t_{s=3} = 9$, $\mbf{y}_1(s) = 1$ for all $s \in \{1,2,3\}$ and 0 otherwise. By the same logic, we obtain $\mbf{y}_2(s) = 1$ for all $s \in \{1,2,\dots,7\}$ and 0 otherwise, resulting in a corresponding $[16 \times 1]$ outcome (label) vector, that can be passed as a tuple $\{\mbf{y}, \mbf{x}\}$ for final stage estimation or learning via classification routines.

\section{Application: Predicting Sleep \& Work Activity from Alternative data using SFCA}

Human time use is an area of research that has drawn enormous interest from many domains of science. Quantifying human sleep patterns is of particular, and pressing concern, with widespread human health problems increasingly associated with so-called `social jet-lag' -- a widespread lack of regular high quality sleep among modern society~\citep{vandeStraat:2015jt} -- leading to calls for a world-wide sleep data collection project~\citep{Roenneberg:2013hc}. The rise of intensive engagement with personal mobile technology, especially near bed-time, has only added to these fears~\citep{Chang:2015jr, Lam:2014eb}.

On the other hand, several studies have shown that human interaction with mobile devices, apps, or the internet more broadly, has the capacity to leave markers of human behaviour in the digital realm~\citep{Walch:2016fa,Quan:2014uh,VisualizingthePuls:2013tf,Althoff:2017hj,Althoff:2017ff} leading to the possibility that despite risks, the increasing digitisation of human life could open the door to remarkable scientific opportunities.

Equivalently, periodic electricity demand signals have long been considered to arise from human behavioral, environmental and economic patterns, with authors applying a wide variety of techniques to predict electricity demand including time-series analysis~\citep{Hyndman:2010co}, functional data analytic methods~\citep{Cabrera:2017be}, markov chain models~\citep{Muratori:2013dc}, neural networks~\citep{An:2013cs} and other digital signals such as mobile (cellular) meta-data~\citep{Bogomolov:2016et}. In each previous case we are aware of, the focus has been to predict electricity demand \emph{from} human behavioral, economic, or environmental signals (or some combination of all three). What appears to have received far less attention is the possibility for electricity demand signals to be used in the opposite sense, i.e. as a \emph{predictor} of human behaviour. It is this latter feature of electricity demand we shall leverage in the present application.

In the following sections of the paper we demonstrate how SFCA can be applied to two separate and novel prediction problems where the object of prediction is the same, namely quantitative sleeping and working phases of human daily behaviour, thus illustrating the power of alternative data to shed light on critical human behavioral patterns, at scale. We first introduce the ATUS dataset from which weighted average sleep start, sleep stop, work start and work stop times for 81 American cities are calculated. Next, we describe the construction of a highly granular internet activity dataset, geo-located to each of the matching ATUS city boundaries, followed by the construction of the electricity demand dataset which we again geo-locate to approximate ATUS spatial boundaries. In both cases, the predictor trajectories so obtained arise from consistently collected, passively measured techno-human signals, which fit into a class of signals that have been recognised for at least a decade as offering immense potential for scientific insight~\citep{Vespignani2009}.

\subsection{ATUS Sleep \& Work Data}
The data used for the prediction of sleep and work is taken from the American Time Use Survey (ATUS). The ATUS is a survey conducted on a sub-sample of the Current Population Survey (CPS). Specific respondents are picked after being surveyed for eight months to have a representative time use allocation by region and demographics. The respondents are interviewed over the phone between 2 to 5 months after CPS has ended to detail all their activities they conducted on the previous day, providing start and stop times for each activity. A prominent activity in the survey is a respondent's sleep and work activity (sleep time, wake time, work begin and work end).

We use \textit{Sleeping} category (ATUS: 010101) as our main input for sleep. Note that the survey provides instructions for coding \textit{Sleeplessness} (ATUS: 010102), which we do not consider. For work we use the categories and its sub categories \textit{Work and Work-Related Activities} (ATUS: 050XXX). The ATUS and CPS data are extracted using an online extraction builder that allows pre-selection of variables \citep{hofferth2013american}. The ATUS and CPS data are merged, as the metropolitan area \textit{fips} code is not readily available in the ATUS data files. 

Given that we are interested in predicting average, city-wide, over-night sleep patterns, we focus on the main portion of over-night sleep, as opposed to `napping' behaviour that may occur for respondents during the afternoon of a given day. Consequently, and following \cite{gibson2018time}, for each respondent, we count the first time diary entry matching the sleeping category after 7 pm as sleep start and the last entry before 12 pm as sleep stop. Every response outside of this time frame was discarded. Similarly, for work we use the first entry of any working category on a day as work start and the last as work stop. The average  start and stop times by metropolitan area is calculated by year using successive difference replicate (SDR) weights provided by the survey. We focus only on metropolitan areas with population exceeding 250,000 people in a given year as the census office warns that estimates with fewer people should be treated with care. We perform this process for sleep and work separately since not every respondent in the survey is working. Finally, this leaves 81 larger metropolitan areas in the US which can be used as input for the wake-work-sleep cycle model. As described below, we match all 81 of these cities for the internet activity data exercise, and 26 for the electricity demand exercise. When comparing prediction performance, we conduct independent analyses by city size, using population categories in all.

\subsection{Internet Activity Data}

\subsubsection{Overview}

Internet activity data were provided by the University of Southern California (USC) PREDICT internet-security database~\citep{Heidemann:2008cv} whilst IP-geolocation information was provided by Digital Element, a highly accurate commercial source~\citep{Gharaibeh:2017,geocompare}. Specifically, we utilised USC PREDICT's IP activity \emph{full census} of all $2^{32}$ IP addresses (NB: since a full scan might take around a month from one source location individual IP measurement time-stamps are spread over the month) as well as 1\% sub-sample scans which provide repeated online/offline observations for clusters of IPs at 11~min intervals~\citep{Heidemann:2008cv}. In each scan, the most basic node-to-node query (a `ping') is sent, asking the target IP if it is presently online, returning a success indicator and return time. In the case that an IP address is not online, or unreachable due to firewalls or other prohibitions, the nearest router or host will respond. In contrast to previous studies such as the `trinocular' approach~\citep{quan2013trinocular}, we make use of every responsive IP address from all types of scans and aggregate the results to a higher spatial scale, such as city, and aggregate the time dimension.

Whilst the fundamental data-generating process is a low dimensional measurement of a single `ping' response, it is the aggregation of these measurements in space and time, coupled to the global scope and consistency of the measurement technology that results in a richly valuable data-source for behavioural inference in the social sciences.

\begin{figure}[ptbh]
    \centering
    \includegraphics[width=0.8\textwidth]{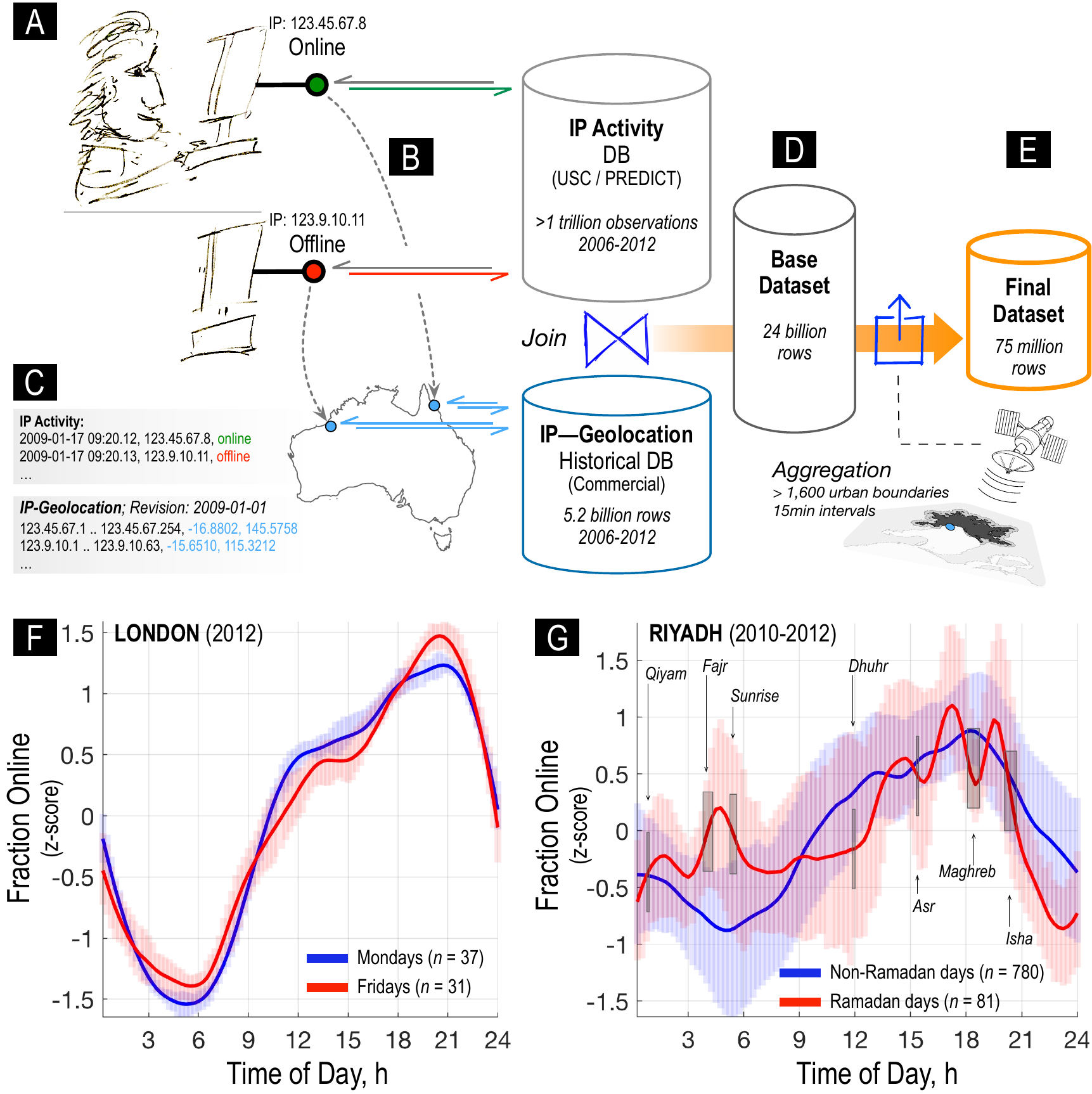}
    \caption{\small Constructing the geo-located, Internet Protocol (IP) activity dataset: \textbf{(a)} Every user on the internet is assigned a unique ID known as an IP address. When a user has an open pathway to the internet it will respond as `online’ when an ICMP probe (ping) is sent by the scan. Any non-response from the IP (e.g. the user’s modem is off or `asleep’, or a firewall is present) will be registered in the IP Activity DB as `offline’. \textbf{(b)} The geolocation (lon, lat) of an IP can be determined by repeated scanning from multiple remote locations. \textbf{(c)} Example IP Activity DB and IP-Geolocation records that can be joined by matching the unique IP to the given IP-range, for the correct time-period. \textbf{(d)} After the join, 24 billion geo-located, IP activity, observations resulted. \textbf{(e)} Finally, the base observations are spatially aggregated using over 1,600 urban boundaries. \textbf{(f)} \textbf{(g)} Remarkably clear fingerprints of human behaviour are revealed in the final intra-diurnal traces for London (the `Friday effect') and Riyadh (the Ramadan effect)}
    \label{figure_overview_data_merge}
\end{figure}

In summary, our method aggregates 75 million rows of online/offline IPs, at 15min intervals, at 1,647 urban-boundaries in 122 countries. Significantly, our data cover a key phase of the internet's global expansion with the user base doubling from roughly 16\% to over 35\% during 2006-2012~\citep{ITU:2016tu}. Figure \ref{figure_overview_data_merge} introduces the main steps in our data preparation pipeline visually. First, the data were collected by pinging individual IPs (A), these measurements were then geo-located (B and C), the resulting tuples were then aggregated to 15 minutes intervals (D), and each 15 min measurement by location was then merged with a city boundary to create a trajectory of internet activity (E). Two example panels are provided of the resulting intra-diurnal traces, demonstrating remarkably clear fingerprints of human behaviour for London (the `Friday effect') and Riyadh (the Ramadan effect) (F and G).

\subsubsection{Details}

\begin{figure}[tbp]
\begin{center}
         \includegraphics[angle=0,width=0.9\textwidth]{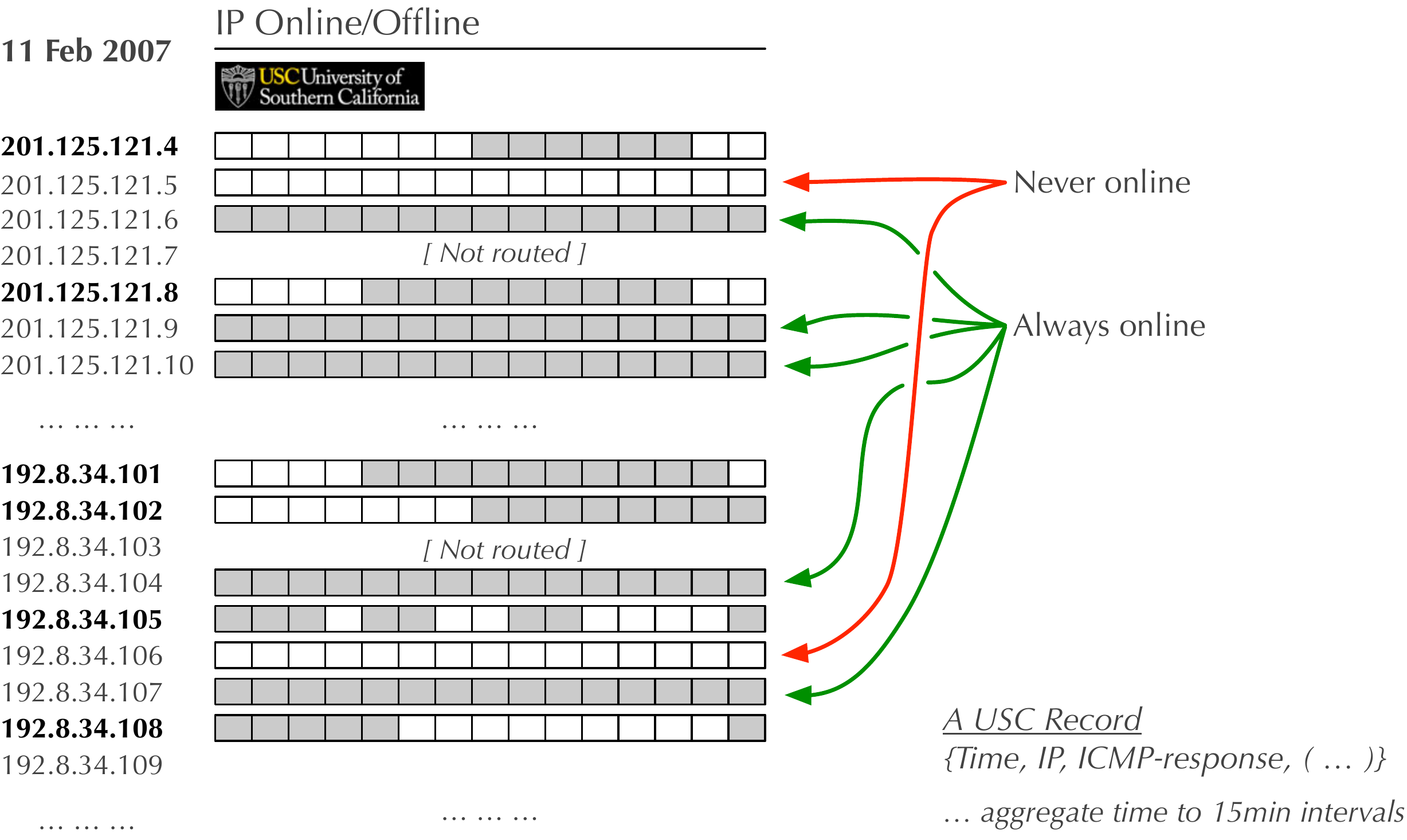}
\caption{Overview of raw Internet Protocol (IP) address activity data. Each record provides a timestamp of measurement, the IP address measured, and the ICMP (ping) response (online or offline), and a response time. All routed IPs (those in use) are measured throughout the day, being found to be never online (white boxes), always online (fully filled grey boxes), or a pattern of online and offline. IPs listed in bold-face text generate the online/offline activity dynamics which are the focus of the sleep and work application in this study.}\label{ip_online_offline}
\end{center}\end{figure}
\begin{figure}[tbp]
\begin{center}
         \includegraphics[angle=0,width=0.9\textwidth]{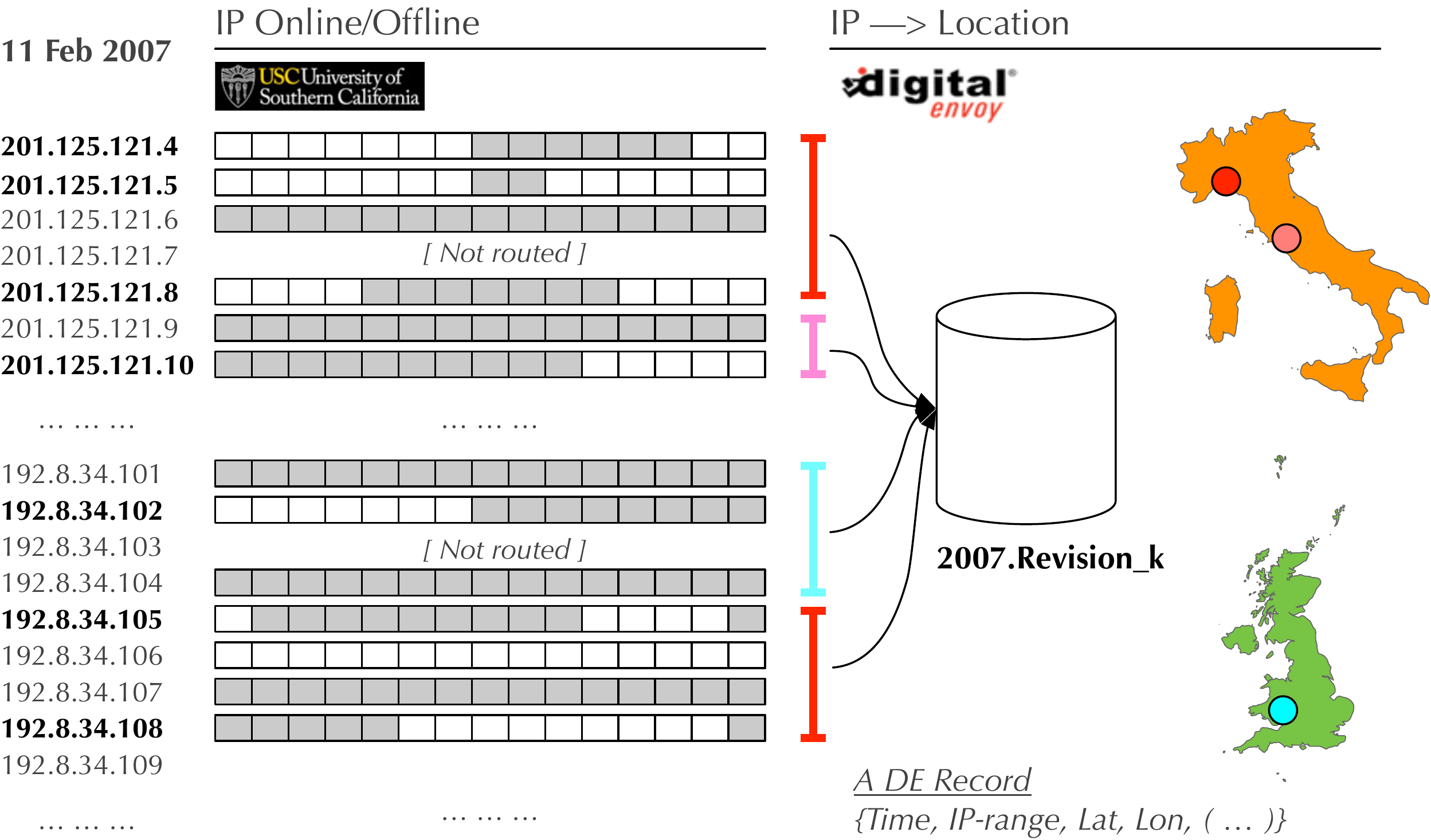}
\caption{The association of internet protocol (IP) addresses with known geospatial locations for IP-ranges. A Digital Envoy (DE) record comprises of a timestamp (typically a bi-monthly revision), an IP-range, and a geo-location in latitude and longitude. The data merge problem is thus defined as associating a given individual IP activity measurement, at a given time, to the correct IP-range, in the appropriate geo-location revision, producing a matched geo-located IP-activity dataset.}\label{ip_geo_match}
\end{center}\end{figure}

In Figures \ref{ip_online_offline} and \ref{ip_geo_match} we present an overview of the online/offline activity data in its basic form, and how these data are associated with a geo location. In both figures, we represent observations on a given IP over a contiguous series of 15 min intervals. If a scan found the IP `online' at any time during the interval, we shade the box dark-grey, whilst `offline' IP responses are shaded white. IPs which were found to be non-routed do not produce a scanning time-series. In Fig.~\ref{ip_online_offline} there are a variety of possible IP response time-series indicated, from `never-online' (no `online' segments), to `always online' (all `online' segments).
The second data source is additionally represented in Fig.~\ref{ip_geo_match}. A given contiguous sequence of IP addresses, an \emph{IP range}, are associated with a particular geo location. For example, we have that the IP range \{{\tt 201.125.121.4 .. 201.125.121.8}\} and the IP range \{{\tt 192.8.34.105 .. 192.8.34.108}\} are jointly associated with the red location (a city in Italy). Importantly, the IP -- geo-location database contains 516 sub-databases, each issued every one to three weeks with the current geo-location (of millions of IP ranges) over entire IPv4 space. The data-join task is consequently complicated by this dynamic IP -- geo-location characteristic. When joining a geo-location to a given IP activity signal, one must find the correct historical record from the IP -- geo-location database to make the join.

With the matched IP -- geo-location database, aggregation at city-years could proceed. City boundaries were spatially defined by the Lincoln Institute of Land and Policy city boundaries \citep{angel2010planet} and used to associate IP -- geo-location measurements to specific cities. To capture daily variation in the online/offline component of internet activity, the fraction of online IPs at a given temporal segment, $s$, $x_{on}(s)$ was calculated from the count of unique online versus offline IPs, $\#^{\text{on}}(s)$, and  $\#^{\text{off}}(s)$ respectively,
\[
x_{on}(s) = \frac{\#^{\text{on}}(s)}{\#^{\text{on}}(s) + \#^{\text{off}}(s)} \, \, .
\]

Next, to account for amplitude variation due to variable always-on server activity in different cities, the daily traces underwent registration to create a synthetic week (Monday .. Sunday) by city and year. Each daily $x_{on}(s)$ trace (96, 15min segments) was first normalised to the [0,1] interval. Next, for each city--year, synthetic weeks were prepared by collecting all days corresponding to each day of the week (e.g. all `Mondays', `Tuesdays', and so on). The average of all such days was taken to generate a representative, contiguous week of seven days in sequence (`Mon', `Tue', ..., `Sun'). Finally, a robust smoother \cite{Garcia:2010hn} was applied to each synthetic week-day to account for any residual noise. A parameter setting of 500 for the smoother was used (a strong setting for this particular procedure).

We use the synthetic week as input to create additional features, which we use as input for model comparison later. To capture the rates of change between the segments we include the first and second differences, and we create dummy variables by weekday, which are 1 at the peak and trough of a day respectively, 0 otherwise. In case important information is held within the full weekly synthetic trace, each contiguous synthetic week by city--year was compressed using wavelet compression (\textit{sym3}, level 7) providing 10 coefficients as further features. Finally, the absolute latitude of each metropolitan area was included as input.

\subsection{Electricity Demand Data}

Electricity demand data for the US were obtained from the Federal Energy Regulatory Commission (FERC) website~\footnote{https://www.ferc.gov/docs-filing/forms/form-714/data.asp}. Specifically, we use `Form 714' filings by US `balancing authorities', at hourly level, for each of 365 days from 1am, 1 Jan 2006, until 31 Dec 2017. Within `Part III - Schedule 2. Planning Area Hourly Demand' balancing authorities must enter, for each hour of the day, the `area's actual hourly demand, in megawatts'.

Since balancing authorities in the US are not aligned to any other geo-spatial boundaries we are aware of, an approximate spatial match to ATUS (MSA) city boundaries was conducted. In all, 26 balancing regions were identified as having comprehensive coverage over a given ATUS city boundary, typically covering at least 99\% of the city region. In five cases, the match was less than 90\%, with two cases matching between 70\% and 80\% of the ATUS city boundary. It should be noted that in the other direction, balancing authority areas are typically much larger than ATUS city boundaries, with some ATUS city boundaries comprising a little over 5\% of the balancing area (e.g. Nevada Power Company supplies a large area of the state, with the city of Las Vegas (Paradise, NV) comprising just 5.4\% of the larger region). However, due to the concentration of people living within major cities within the balancing region, it is reasonable to assume that the balancing authority region's power demand is driven, in large part, by the population centres within the larger boundary.

\begin{figure}
    \centering
    \includegraphics[width=0.8\textwidth]{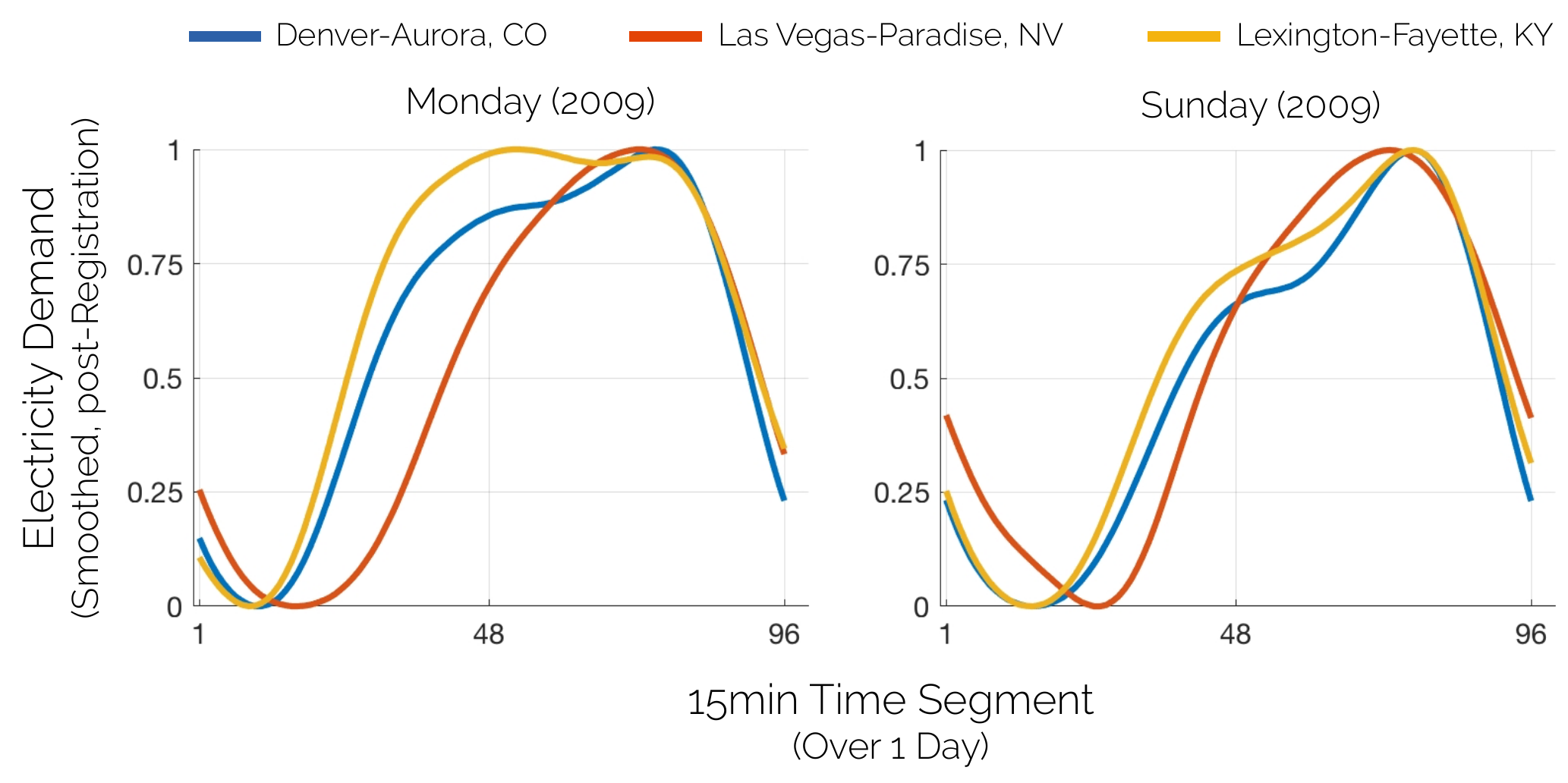}
    \caption{Electricity demand trajectories for three ATUS-matched balancing authorities during 2009. Traces shown are resultant from pre-processing and registration of average Monday (left) and Sunday (right) demand.}
    \label{fig:electricity}
\end{figure}

To prepare the electricity demand data for processing, a number of steps were undertaken. After ingestion and matching to ATUS city boundary areas, each date was assigned a day of the week to facilitate aggregation to average day of the week (dow) predictor trajectories. Prior to aggregation, hourly data were lightly smoothed and then down-scaled to 15min segments using spline interpolation. Following this, day of the week average trajectories per calendar year were obtained by taking the average demand per 96, 15min segment, and then registration by normalising the resultant trace to range $[0,1]$ (see Fig.~\ref{fig:electricity}). Feature engineering proceeded by following similar steps to internet activity trajectories with day of week peak and trough segments coded to 1 and 0 otherwise, first and second differences of normalised demand were obtained (with loss of the first two segments per day), and finally, wavelet compression coefficients (6 per day of the week) were generated and stored as static features by city--year--dow.

\subsection{Applying SFCA to the Sleep/Work Problem}

The Sleep/Work prediction from internet activity or electricity demand problem fits within the inclusion criteria for SFCA. Sleep or work start/stop outcomes are given on the time-of-day domain ($\mbf{Y}$), corresponding to equivalent time-of-day measurements, either at 15min or hourly ($\mbf{T}$), for internet activity and electricity demand predictor trajectories. As such, $\mbf{Y} \subseteq \mbf{T}$ is satisfied and SFCA can be applied to the problem.

\begin{figure}
    \centering
    \includegraphics[width=0.9\textwidth]{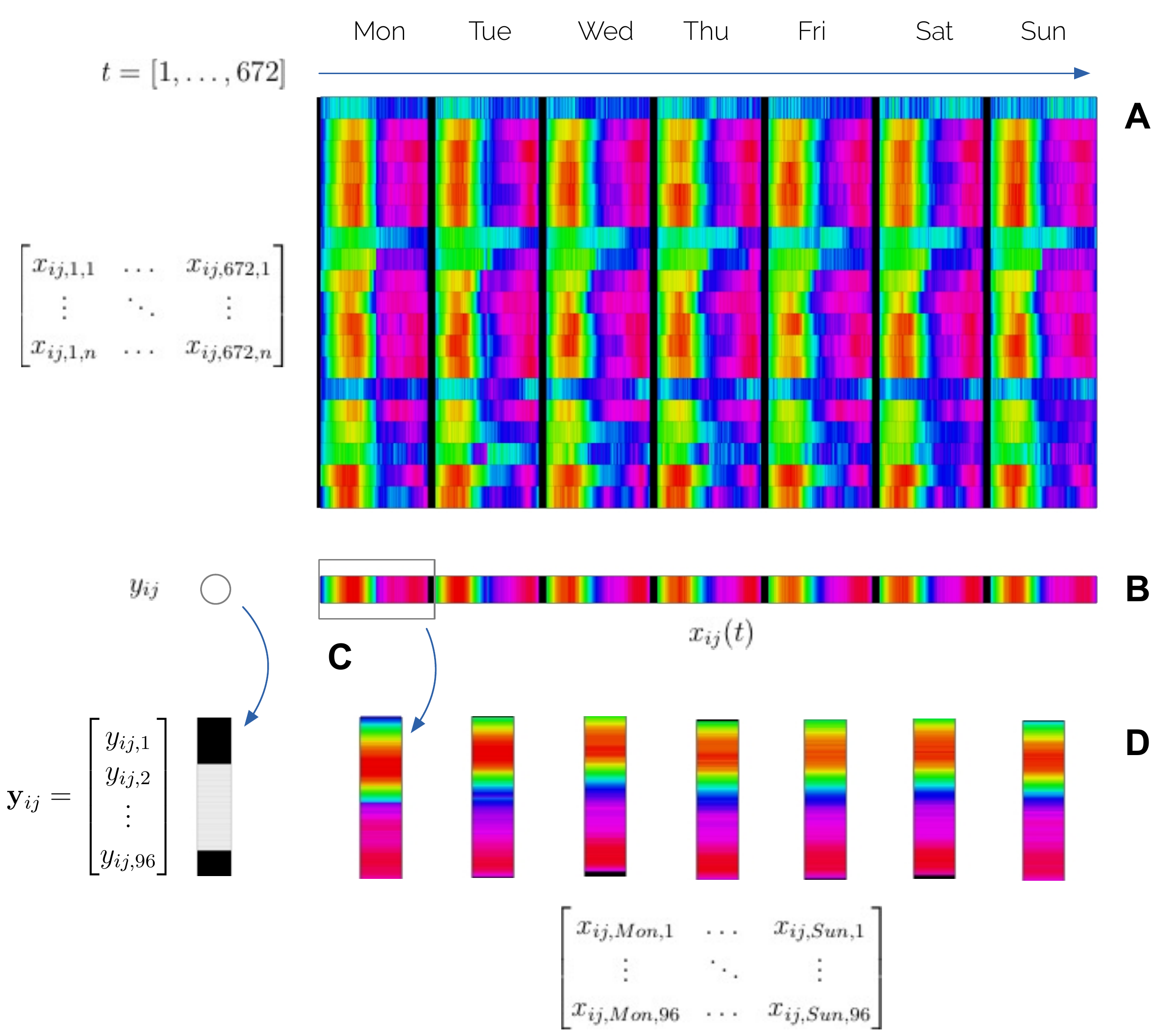}
    \caption{Segmented Functional Classification Analysis (SFCA) overview: A) a matrix of $n$ weekly observations of predictor $x_{ij}$, each at 96 sequential, 15min segments of the days Mon..Sun (672 observations in each record); B) observations are collapsed by averaging (at segments) to a representative functional trace $x_{ij}(t)$, which together with the scalar outcome $y_{ij}$ are ready for functional regression analysis; C) alternatively, in this study, both the outcome and the predictors are transformed by packing (or `stacking') enabling the application of classification analysis.}
    \label{fig:schem}
\end{figure}

\paragraph{Pre-processing}: For a given geo-spatial unit (e.g. city), $i$ and in year, $j$ we have $n$ weekly observations on internet activity or electricity demand. For the former, observations are given over 96 regular, and complete 15min intervals per day, and day-of-the-week is used as a predictor dimension, i.e. $t_1 = [\textrm{Mon 00:15},\textrm{Mon 00:30},\dots,\textrm{Sun 23:30}, \textrm{Sun 23:45}]$ (672 intervals in all, Fig.~\ref{fig:schem} A),
\[
X_{ij}(t) =
\begin{bmatrix}
          x_{ij,1,1} & \dots & x_{ij,672,1} \\
           \vdots & \ddots & \vdots \\
           x_{ij,1,n} & \dots & x_{ij,672,n} \\
\end{bmatrix}\, .
\]
For the latter, observations are given over 24 hourly intervals and again day-of-the-week is used as a predictor dimension, i.e. $t_2 = [\textrm{Mon 00:00},\textrm{Mon 01:00},\dots,\textrm{Sun 22:00}, \textrm{Sun 23:00}]$ (168 intervals in all). In order that electricity demand data can be processed in an analogous setting to the internet activity data, the electricity demand trajectories were smoothed then down-scaled by spline interpolation to 15min intervals as described earlier.

Since outcomes for each city are given as an annual average, all Mondays, Tuesdays, Wednesdays (and so on) are averaged to create average predictor trajectories for each day of the week at a given city, $x_{ij}(t)$ (Fig.~\ref{fig:schem} B). Taken together, segmentation is naturally defined for these data over 96 15min intervals per day, with segment $s=1$ corresponding to the first 15min of the day, and segment $s=96$ to the last.

\paragraph{Stacking}: Next, we apply the stacking transformation to the average signal at each city (Fig.~\ref{fig:schem} C)  $x_{ij}(t)$ by taking the transpose of each 96 segment, 24h period sub-set of $x_{ij}(t)$, to create the object,
\[
\mathbf{x}_{ij} =
\begin{bmatrix}
          x_{ij,Mon,1} & \dots & x_{ij,Sun,1} \\
           \vdots & \ddots & \vdots \\
           x_{ij,Mon,96} & \dots & x_{ij,Sun,96} \\
\end{bmatrix}\, .
\]
In terms of dimensionality, stacking has transformed the problem, at each city--year observation, from a $[1 \times 672]$ wide problem, to a $[96 \times 7]$ tall problem.

\paragraph{Threshold \& Balance}: In this application, we can utilise the fact that two outcomes for each ATUS activity, both defined on the same domain (e.g. sleep-start / sleep-stop) can be utilised to define the boolean classification object $\mbf{y}_{ij}$.  Namely, define, for each city--year, the activity pair of start ($0\rightarrow 1$ transition) and stop ($1\rightarrow 0$ transition) transition times respectively as $y^{01}_{ij}$ and $y^{10}_{ij}$. A balanced logical outcome vector can then be formed by thresholding, for example, in the case of activity `sleep' which typically occurs in two segments at start and end of the 96 segment day,
\begin{equation}\label{eq:sleep}
\mathbf{y}_{ij}
= \begin{bmatrix}
           y_{ij,1} \\
           y_{ij,2} \\
           \vdots \\
           y_{ij,96}
\end{bmatrix}
= 
\begin{cases}
  1 & \text{for} \quad t_s \leq y^{10}_{ij} \,\, \textrm{or} \,\, t_s \geq  y^{01}_{ij} \\
  0 & \text{otherwise}.
\end{cases}
\end{equation}
In the case of activity `work', which typically occurs in the middle of the 96 segment day, the class 1 condition of \eqref{eq:sleep} is adapted from `or' to `and'. For example, for the outcome `sleep', suppose that a given city--year had average sleep start and sleep stop times of 22:15 and 06:47 respectively $\mathbf{y}_{ij}$ will take the value of class 1 for segments $\{1,\dots,27\}$ and $\{90,\dots,96\}$. In both cases, since sleep duration or work duration roughly covers 6-10 hours of a given day, whilst not perfectly balanced, the outcome object is sufficiently balanced for classification.

\subsection{Performance Evaluation}
We benchmark out proposed methodology against well established methods in functional data analysis, regression prediction and classification analysis. The classification methods use our stacked input matrices after processing as outlined above. To allow a comparison of the final output from the SFCA methods with the regression methods, we undertake a conversion to a continuous prediction by decomposition, described below.

\subsubsection{Continuous Prediction from SFCA}

\begin{figure}
    \centering
    \includegraphics[width=0.8\textwidth]{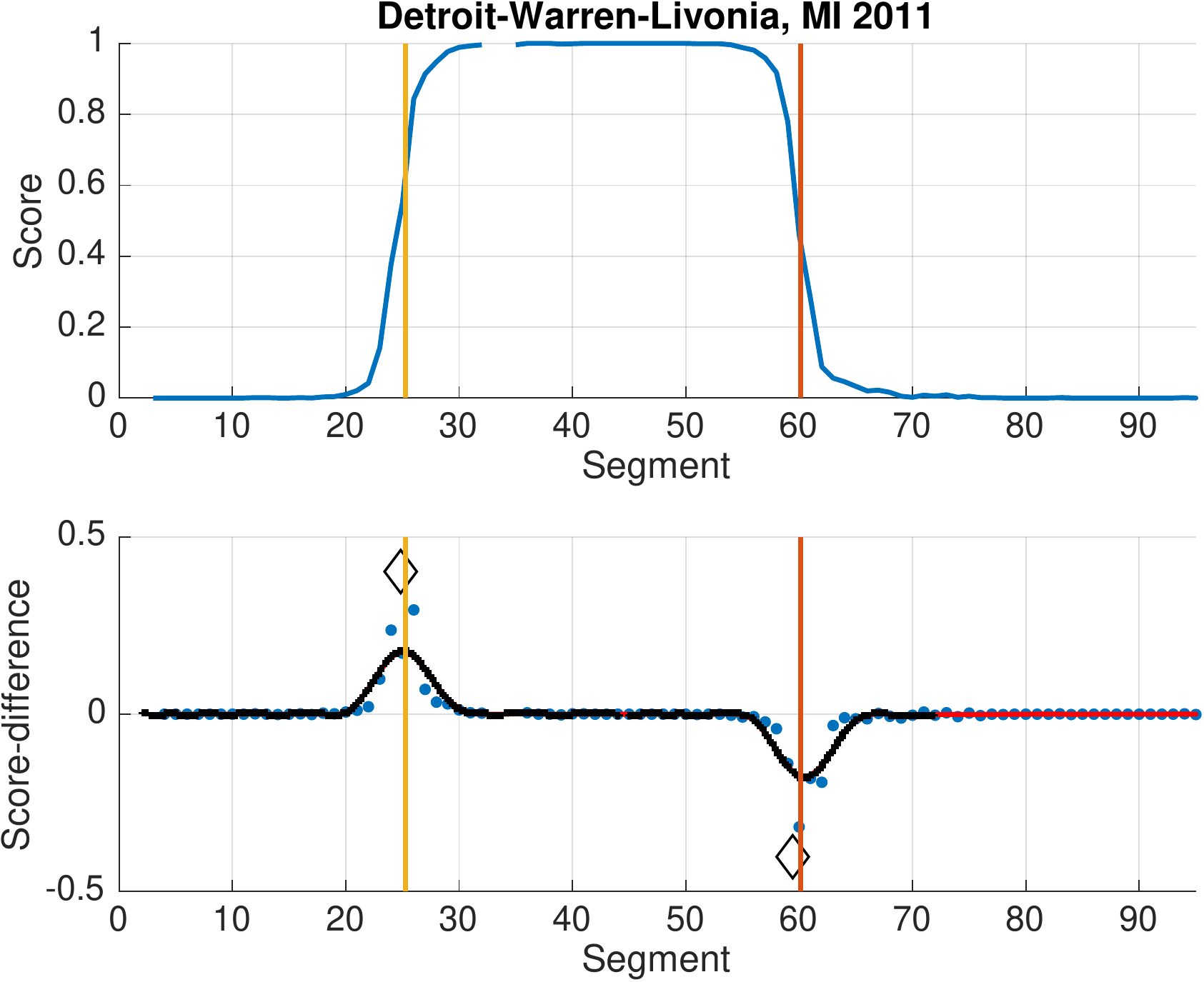}
    \caption{Converting SFCA predictions back to continuous time. (top) Trained model prediction scores for sleep-start (left) and sleep-end (right) are shown in blue. (bottom) Prediction-score differences (blue markers) underpin wavelet de-noised and interpolated signals (black markers) to identify the continuous time switching moment to- or from- sleep. Diamonds show the final sleep-switch prediction location (platted with a y-shift up/down to avoid over-plotting). In both panels, the yellow and red vertical lines show the ground-truth average sleep-start and sleep-stop times for this city-year.}
    \label{fig:warren}
\end{figure}

Using $n$-fold cross-validated prediction, we predict, for each method, each city--year combination based on all the other cities as input. As an example, figure~\ref{fig:warren} displays the independent segment predictions of sleep for the metropolitan area Detroit-Warren-Livonia, Michigan in 2011. The segments are shifted with the day starting at 4pm (Segment 64). The gap in the scoreline are the two segments after midnight, which are not used due the first and second difference of the fraction online input data.

The conversion to a continuous prediction follows the following procedure. First, the shifted prediction score is smoothed slightly with a parameter value of $0.06$~\citep{Garcia:2010hn}, to remove the sharp edges of the independently drawn prediction as well as to interpolate the time gap after midnight. Shifting of the signal is of particular importance to accurately allow the smoothing algorithm to interpolate the gap. Second, the first difference of the smoothed signal is calculated as highlighted by the blue dots in the bottom panel of figure \ref{fig:warren}. Third, wavelet de-noising (\textit{sym8}, level 1) was applied to remove noise but at the same time preserve the sharp gradient from the $0$ to $1$ switch. Fourth, spline interpolation (black line) for the waking and to-sleep sections was applied, separated at 3 am.  Finally, the x-position of the minimum and maximum of the two phases was identified to derive the predicted start and stop times in minutes. The vertical lines represent the survey based (groundtruth) start and stop estimates, while the diamonds mark our prediction outcome (vertical shift of these markers is to avoid over-plotting only).

\subsubsection{Comparison Framework}

To explore the potential advantage of the SFCA methodology proposed herein we undertake a wide-ranging prediction activity where linear and non-linear regression technologies are applied to un-transformed input trajectories alongside classification technologies applied in a SFCA manner. We consider 12 independent prediction exercises for each specific technology: start time, stop time and duration, for sleep and work time-use activities, with input signals from internet or electricity data (3 x 2 x 2).

Each method was evaluated based on their out-of-sample performance on a universe of up to either 81 (internet data) or 26 (electricity data) matched cities with an atomic observation being a city--year. We perform leave-one-out cross-validation (LOOCV) across cities. This evaluation was chosen to represent our use case of obtaining additional insights about a city whose internet or electricity data are available, but no time-use survey data was collected.

To further test methodologies against more or less noisy contexts, we also iterated over five city population groups as follows: $>$ 250,000; $>$ 500,000; $>$ 1,000,000; $>$ 2,500,000; and $>$ 5,000,000. The increase in population size of a city decreases the standard error estimates from the ATUS survey for a given city-year. By applying these filters, we generate 5 independent LOOCV experiments for each data-set with the data basis ranging from 212 city-years for the most inclusive filter with electricity data, to 22 city-years for the least, or accordingly, 825 city-years and 42 city-years for the internet dataset.

To present a single, and exacting, summary statistic for a given methodology applied to a given data type and population filter setting, we calculate the geometric mean ($GM$) of root mean square errors (rmse), in minutes, over the 5 population filter LOOCV experiments for each method.

Our comparison included the following methods (with labels used in subsequent result reporting).

\paragraph{Regression Methods} For regression methods we compare standard ordinary least squares (\verb=ols=), shrinkage methods with lasso or ridge penalties~\citep{tibshirani1996regression} (\verb=lasso=, \verb=ridge=), tree based methods with bagging~\citep{breiman2001random} (\verb=r-tree=, \verb=r-tree(bg)=) or boosting~\citep{chen2016xgboost} (\verb=r-tree(bs)=). In addition, we also test model variants where we put more weight on cities with a higher number of survey respondents (\verb=lasso(w)=, \verb=r-tree(bg)(w)=, \verb=r-tree(bs)(w)=).

\paragraph{Functional Data Analysis Methods} Here we use the well established functional linear regression~\citep{Yao:2005he} (\verb=flr=) and the functional additive model (FAM)~\citep{muller2008functional} (\verb=fam=).

\paragraph{SFCA Classification Methods}  SFCA enabled classification methods considered included Support Vector Machine (SVM) (\verb=svm=), classification trees with bagging (\verb=c-tree(bg)=) and boosting (\verb=c-tree(bs)=) and a logistic regression model with ridge (\verb=logr(ridge)=) and lasso (\verb=logr(lasso)=) penalties~\citep{hastie2009elements}. As with regression, we add model variants with weighted observations (\verb=c-tree(bg)(w)=, \verb=c-tree(bs)(w)=).

\subsection{Results}

In Table~\ref{tb:ipresults} and \ref{tb:elresults} we present performance comparison results of the application of SFCA to the ATUS sleep/work prediction exercise from internet activity and electricity demand respectively. The top panel of each table shows results from methods that leverage regression on un-transformed/un-segmented data, whilst the bottom panel provides results from methods which leverage SFCA. Results in each panel are sorted in ascending order by the $GM(RMSE)$ of the sleep:start problem. In addition, we provide an indicator of the best performing un-transformed prediction method, and bold-face for the best method overall. Underline indicates any SFCA method which out-performs the best non-SFCA method.

\begin{table}[htbp]
\footnotesize
\centering
\caption{\footnotesize Internet Data to Sleep/Work - Performance Comparison (81 US Cities, 2006-2012)}\label{tb:ipresults}
\begin{tabular}{llclllllll}
\hline
		    & 		    & 		 & \multicolumn{7}{c}{$GM(RMSE)$ (min)} \\
\cline{4-10}
		    & 		    & 		 & \multicolumn{3}{c}{Sleep} && \multicolumn{3}{c}{Work} \\
\cline{4-6}
\cline{8-10}
Method Name & Type & $GM_n$ & Start & Stop & Duration && Start & Stop & Duration \\
\hline
r-tree(bg)  & REG-T & 5 & 15.19\^{}  & 25.84\^{} & 24.00\^{} && 41.53\^{} & 45.81 & 49.51\\
lasso &  pREG & 5 & 15.47  & 25.87 & 23.83 && 41.90 & 45.10 & 63.41\\
ridge &  pREG & 5 & 15.52  & 25.93 & 25.09 && 41.97 & 45.07\^{} & 48.55\^{}\\
flr &  FDA & 5 & 15.59 &  26.38 & 24.13 && 42.98 & 46.66 & 50.30\\
fam &  FDA & 5 & 15.78 &  27.42 & 24.75 && 44.90 & 55.67 & 53.07\\
r-tree(bg)(w) & REG-T & 5 & 16.89 & 29.18 & 27.23 && 50.22 & 54.02 & 58.14 \\
lasso(w) & pREG & 5 & 17.26 & 33.28 & 27.59 && 50.77 & 53.43 & 101.03\\
r-tree(bs) &  REG-T & 5 & 17.85  & 30.81 & 29.28 && 49.38 & 54.73 & 59.46\\
r-tree(bs)(w)  & REG-T & 5 & 20.03  & 34.89 & 33.42 && 61.22 & 64.36 & 70.48\\
r-tree  & REG-T & 5 & 20.54  & 35.87 & 31.15 && 52.29 & 61.37 & 64.77\\
ols  & REG & 5 & 91.60  & 59.40 & 60.35 && 90.12 & 109.40 & 93.29\\
\hline
c-tree(bg)  & SFCA-T & 5 & \bf \ul{9.64}  & \bf \ul{18.45} & \bf \ul{19.48} && \ul{16.33} & \bf \ul{17.97} & \bf \ul{28.48}\\
c-tree(bg)(w)  & SFCA-T & 5 & \ul{11.39}  & \ul{21.75} & 27.38 && \bf \ul{15.49} & \ul{20.60} & \ul{31.12}\\
c-tree(bs)  & SFCA-T & 5 & 26.26  & 30.59 & 39.32 && \ul{41.24} & 54.27 & 70.25\\
c-tree(bs)(w)  & SFCA-T & 5 & 27.21  & 34.87 & 43.47 && 43.89 & 65.27 & 78.89\\
logr(ridge)  & SFCA-pREG & 5 & 108.74  & 81.04 & 113.71 && 127.05 & 177.00 & 235.86\\
svm  & SFCA-SVM & 5 & 139.97  & 65.98 & 159.97 && 141.74 & 160.36 & 171.70 \\
logr(lasso)  & SFCA-pREG & 5 & 186.53  & 99.41 & 219.80 && 171.39 & 218.79 & 312.12\\
\hline
\multicolumn{10}{p{0.9\textwidth}}{\it Notes:
(\^{}), best performance under regression methods; underlined text, SFCA performance better than best performing regression method; boldface text, best performance under any method.
For details of methods, refer to text.
}\\
\end{tabular}
\end{table}

\begin{table}[htbp]
\footnotesize
\centering
\caption{\footnotesize Electricity Data to Sleep/Work - Performance Comparison (26 US Cities, 2006-2017)}\label{tb:elresults}
\begin{tabular}{llclllllll}
\hline
		    & 		    & 		 & \multicolumn{7}{c}{$GM(RMSE)$ (min)} \\
\cline{4-10}
		    & 		    & 		 & \multicolumn{3}{c}{Sleep} && \multicolumn{3}{c}{Work} \\
\cline{4-6}
\cline{8-10}
Method Name	&	Type	& $GM_n$ & Start & Stop & Duration && Start & Stop & Duration \\
\hline
flr			& FDA & 5 & \bf 13.53\^{} & 23.36 & 22.92 && 41.26 & 49.37 & 52.76 \\
r-tree(bg)	& REG-T &  5 & 13.73 & 23.33 & 22.14\^{} && 39.41\^{} & 47.76 & 48.39\^{} \\
fam 		& FDA & 5 & 14.18 & 22.67\^{} & 23.15 && 42.27 & 51.71 & 55.09 \\
lasso 		& pREG & 5 & 15.13 & 23.12 & 22.26 && 40.41 & 46.97\^{} & 49.23 \\
ridge 		& REG-T & 5 & 15.13 & 23.42 & 22.44 && 40.36 & 47.19 & 49.56 \\
r-tree(bg)(w)  & REG-T & 5 & 15.72 & 27.20 & 25.15 && 51.88 & 56.68 & 54.68 \\
r-tree 		& REG-T & 5 & 16.48 & 28.62 & 28.38 && 54.49 & 60.37 & 62.03 \\
r-tree(bs)  & REG-T & 5 & 16.59 & 29.74 & 27.92 && 49.09 & 56.48 & 58.01 \\
lasso(w) 	& pREG & 5 & 16.71 & 25.97 & 25.17 && 47.65 & 55.48 & 55.45 \\
r-tree(bs)(w) & REG-T & 5 & 18.49 & 31.29 & 29.62 && 61.48 & 68.75 & 61.33 \\
ols			 & REG & 5 & 56.58 & 64.84 & 64.09 && 77.27 & 99.55 & 117.65 \\
\hline
c-tree(bg)   & SFCA-T & 5 & 14.13 & 23.80 & 23.83 && \bf\ul{22.91} & \ul{31.16} & \ul{43.31} \\
c-tree(bs)(w)& SFCA-T & 5 & 14.26 & 23.66 & 24.56 && \ul{28.84} & \ul{39.03} & \ul{45.94} \\
c-tree(bs)   & SFCA-T & 5 & 14.76 & \bf\ul{21.21} & \bf\ul{21.41} && \ul{22.95} & \bf\ul{28.31} & \bf\ul{35.69} \\
c-tree(bg)(w)& SFCA-T & 5 & 14.91 & 25.77 & 28.13 && \ul{24.23} & \ul{36.07} & 49.12 \\
logr(ridge)  & SFCA-pREG & 5 & 45.46 & 57.47 & 34.28 && 83.43 & 129.03 & 154.19 \\
svm 		 & SFCA-SVM & 5 & 55.88 & 39.46 & 46.93 && 87.30 & 113.54 & 123.55 \\
logr(lasso)  & SFCA-pREG & 5 & 79.41 & 92.66 & 53.81 && 107.13 & 308.12 & 292.73 \\
\hline
\multicolumn{10}{p{0.9\textwidth}}{\it Notes:
(\^{}), best performance under regression methods; underlined text, SFCA performance better than best performing regression method; boldface text, best performance under any method.
For details of methods, refer to text.
}\\
\end{tabular}
\end{table}

We find that across the 12 problem environments, representing in all, 60 independent modelling exercises (five each for the different population filters), SFCA methods out-perform non-SFCA methods in all but one environment (electricity:sleep:start). In this particular case, the margin of difference is small: 36s in around 14min (840s), or less than 5\%. Indeed, in the case of electricity demand (Table~\ref{tb:elresults}), for sleep prediction, the performance is roughly similar between SFCA and non-SFCA methods, with margins of gain at most 6.4\% (electricity:sleep:stop), but for work prediction, the margins of gain for SFCA over non-SFCA methods is large, ranging from 26\% to 42\%. Similar gains are found across all prediction environments with internet activity (Table~\ref{tb:ipresults}) where performance gains for SFCA over non-SFCA methods range from 19\% (internet:sleep:duration) to 60\% (internet:work:stop).

We find that across all problem domains, SFCA-T (tree type) methodologies were the most powerful with honours being roughly equally split between bagged and boosted methods. This gives weight to the idea that the potency of SFCA lies in the ability of classification methodologies to identify specific boundaries in segmented data. A point we return to below. For non-SFCA methods, there was more of a mix of high performing methodologies, including bagged regression trees, penalised (ridge or lasso) regression, and both variants of FDA (functional linear regression, functional additive model).

\begin{figure}[htbp]
    \centering
    \includegraphics[width=\textwidth]{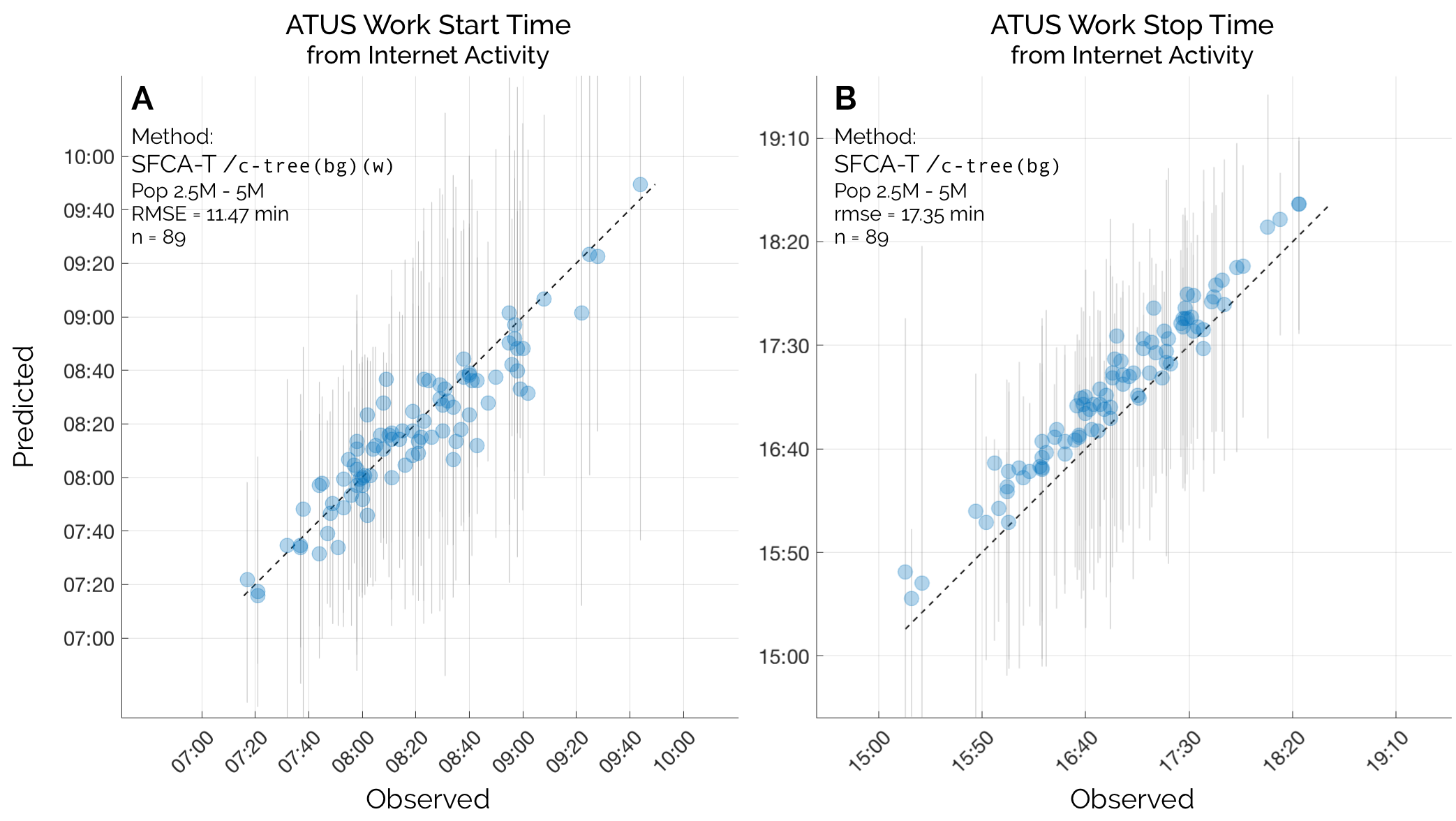}
    \caption{Example Internet activity data to ATUS Work start (panel A) and stop (panel B) predicted vs. observed: best fitting models applied to 89 matching city-years with population from 2.5 to 5 million. Grey bars indicate 95\% confidence interval for observed data; all predicted points fall within respective intervals.}\label{fig:scatter_internet}
\end{figure}
\begin{figure}[htbp]
    \centering
    \includegraphics[width=\textwidth]{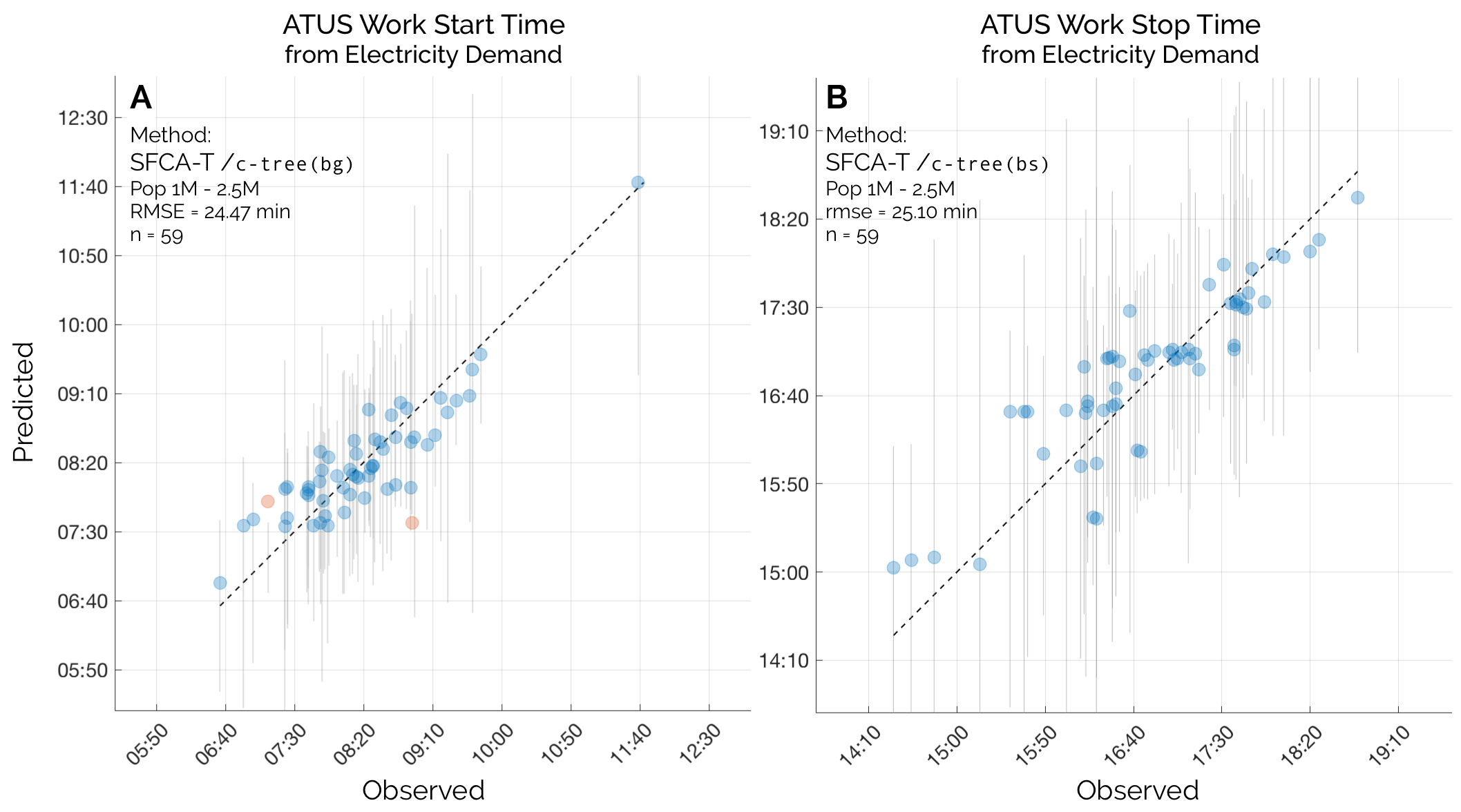}
    \caption{Example electricity demand data to ATUS Work start (panel A) and stop (panel B) predicted vs. observed: best fitting models applied to 59 matching city-years with population from 1 to 2.5 million. Grey bars indicate 95\% confidence interval for observed data; all predicted points fall within respective intervals, excepting two predictions for work start (panel A), both due to the Nashville-Davidson--Murfresboro--Franklin, TN region.}\label{fig:scatter_electricity}
\end{figure}

In Figures~\ref{fig:scatter_internet} and \ref{fig:scatter_electricity} we provide example scatter plots of out of sample activity time predictions versus observed values for work start and stop times, using internet activity and electricity demand data respectively. For the internet plots, we show results from 89 city-years for the 2.5-5.0 million population city filter. For electricity data, given the smaller number of matched city-years, we provide results for 59 city-years from the larger 1-2.5 million population city filter.  In all plots we show predictions from the best performing model versus observed ground-truth averages for each city. In addition, we present 95\% confidence intervals for the ATUS average work start/stop times for each city as vertical lines in the predicted dimension, enabling quick verification of whether a prediction falls inside a statistically acceptable range. Where a prediction lies outside this range, we colour the point orange, with blue markers indicating acceptable variation.

As shown, SFCA-T like model predictions perform very well in all four domains at this city size with RMSE less than 18min for predictions from internet activity data, and around 25min for electricity demand data. Indeed, acceptable predictions (within confidence intervals) are found across both internet derived panels (Fig.~\ref{fig:scatter_internet}), and in all but two city-years in the two electricity demand panels (Fig.~\ref{fig:scatter_electricity}). The outstanding city-years in Fig.~\ref{fig:scatter_electricity}(A) are both due to city years from the Nashville--Murfreesboro--Franklin control region, Tennessee. The two years in question arise from near boundary cases over 12 city-years of work:start data for the region: the earliest for this region at 7:10am (in 2007); and the second-latest at 8:55am (2011). Nevertheless, despite these two out of bounds predictions, the scatter plots demonstrate, across 296 point comparisons, the wide range of applicability of SFCA-T methods to predict human activities with observed average times varying by up to five hours across city-years.

\section{Discussion}

In summary, the present study makes two contributions: first, by introducing a new functional data-set of internet activity use worldwide and applying this to the problem of predicting human intra-diurnal time use patterns; and second, by presenting a new methodology, SFCA, that is shown to out-perform a battery of existing methodologies from across the functional-, linear- and machine learning- toolsets.


On the first -- a new granular dataset on internet activity, our work relates to a small, but rapidly growing number of studies that have demonstrated the utility of passively collected `big data' for the pursuit of quantitative social science. This alternative data were not initially collected for social data-science purposes, instead, the data arose as a by-product of the particular service delivery in question. Eventually, it has been creatively applied to long-standing (and at times, entirely novel) social science research questions.  Nevertheless, none of the existing data sources offer the unique attributes of geo-located internet activity data as developed and analysed here. Geo-located data on cell-phone meta-data, internet activity or online user engagement requires individual agreements with private corporations making it very challenging to construct a dataset that has global coverage and contains consistent measurements. 


In addition, app user data provide an engagement signature bounded by the app's user base on the one hand, and the in-app time-use profile on the other, making a one-to-one mapping between the universe of internet use traces and the given app's use-traces unlikely. Satellite imagery is perhaps the most closely related source to geo-located internet activity data as it shares the features of global scope, consistency of collection, and low-dimensionality (night-time luminosity of 1 square kilometre sections of the Earth's surface), however, the temporal resolution of night-time luminosity data is at most daily with publicly available sources aggregated to annual readings. As such, for intra-diurnal human activity analysis as reported here, satellite imagery is an unlikely candidate.

These considerations simultaneously highlight both the complimentary nature of geo-located internet activity data and its uniqueness. As demonstrated by the predictive performance of internet activity data for human activity as reported in this work, low dimensionality does not prevent the leverage of the unprecedented scope and granularity of global IP scan data for scientific inquiry.  Furthermore, it is worth noting that online/offline scan data is unique in one, further, and particularly striking characteristic: with sufficient know-how, the data may be gathered by \emph{anyone} who is online -- no government agency, nor private company, nor intermediary of any kind is required to enable the collection of low-dimensional, anonymous IP online/offline data. For this reason, for so long as the internet retains its democratic foundations, we consider online/offline data of the kind employed here as fundamentally `open'.

On the second contribution -- a new highly effective modelling methodology we have called \emph{segmented functional classification analysis} (SFCA), we provide some reflections on the likely reasons for its strong performance across two functional datasets and six independent prediction problems. As mentioned above, a key challenge facing any statistical scientist who wishes to build a prediction model of the kind $f : \mathbf{x} \mapsto \mathbf{y}$ is the dimensionality of the problem at hand. For all non-SFCA methodologies considered in the present study, the most restrictive problem environment was to estimate $f$ from the electricity dataset with the most conservative population filter ($>$ 5M). Here, just 22 city--year trajectories were available for estimation, indeed, under LOOCV, model estimation could leverage just 21 rows of functional-like data, with validation on the excluded row. In comparison, by employing 15min- segmentation and then the stack (pack/wide-to-tall) algorithm as part of the SFCA procedure, the problem is transformed to a $21\times94 = 1974$ row environment (recall, 2 segments are dropped due to differencing); a dramatic, two-order, increase in dimensionality.

However, it is important to stress that, in our results, the dimensionality gain of the SFCA methodology does not appear to constitute the totality of its performance improvement. Indeed, careful inspection of Tables~\ref{tb:ipresults} and \ref{tb:elresults} show that, after transformation, binary, penalised regression methods (e.g. \verb=logr{ridge}= and \verb=logr(lasso)=) performed universally worse on the segmented and transformed data than on their un-transformed linear counterparts. Inspection of the results indicates a performance drop of around a factor of two. Instead, with segmented, transformed and binary encoded outcome variables, classification tree models delivered dramatic performance benefits, across internet activity and electricity demand data, especially on the work time-use outcomes. As such, it would seem that the real performance benefit of SFCA is due to its ability to transform functional-data like problems into classification problems, at once scaling up dimensionality \emph{and} porting the problem type to work with modern statistical machine learning tools in classification mode.

Beyond the present applications, it will be of interest to extend SFCA in at least two dimensions. First, applying SFCA to other problem domains, beyond human intra-diurnal time use prediction, where such problems satisfy the SFCA inclusion criteria. Second, by expanding the technology set that SFCA makes available to the statistical scientist, for example, to deep learning methodologies which thrive in tall-scale problem environments.


%



\end{document}